\documentclass[english,preprint,1p]{elsarticle}
\usepackage{bm}
\usepackage{amsmath}
\usepackage{amssymb}
\usepackage{graphicx}
\usepackage{esint}

\makeatletter

\biboptions{sort&compress}

\usepackage{bm}
\usepackage{dcolumn}
\usepackage{amsthm}
\usepackage{babel}
\usepackage{lineno}

\journal{Journal of Computational Physics}

\makeatother

\usepackage{babel}
\begin{document}
\begin{frontmatter}



\title{Canonical symplectic structure and structure-preserving geometric
algorithms for Schr\"{o}dinger-Maxwell systems}


\author[ustc,leetc]{Qiang Chen\corref{cor1}}

\ead{cq0405@ustc.edu.cn}

\author[ustc,pppl]{Hong Qin\corref{cor2}}

\ead{hongqin@ustc.edu.cn}

\author[ustc]{Jian Liu}

\author[ustc]{Jianyuan Xiao}

\author[ustc]{Ruili Zhang}

\author[ustb]{Yang He}

\author[ustc]{Yulei Wang}

\cortext[cor1]{Corresponding author}

\cortext[cor2]{Principal Corresponding author}

\address[ustc]{School of Nuclear Science and Technology and Department of Modern
Physics, University of Science and Technology of China, Hefei, Anhui
230026, China}

\address[leetc]{Luoyang Electronic Equipment Testing Center, Luoyang 471000, China}

\address[pppl]{Plasma Physics Laboratory, Princeton University, Princeton, NJ 08543,
USA}

\address[ustb]{School of Mathematics and Physics, University of Science and Technology
Beijing, Beijing 100083, China}
\begin{abstract}
An infinite dimensional canonical symplectic structure and structure-preserving
geometric algorithms are developed for the photon-matter interactions
described by the Schr\"{o}dinger-Maxwell equations. The algorithms
preserve the symplectic structure of the system and the unitary nature
of the wavefunctions, and bound the energy error of the simulation
for all time-steps. This new numerical capability enables us to carry
out first-principle based simulation study of important photon-matter
interactions, such as the high harmonic generation and stabilization
of ionization, with long-term accuracy and fidelity. 
\end{abstract}
\begin{keyword}
Schr\"{o}dinger-Maxwell equations \sep symplectic structure \sep
discrete Poisson bracket \sep geometric algorithms \sep first-principle
simulation 
\end{keyword}
\end{frontmatter}



\section{Introduction}

\label{sec:1}

Started with the photoelectric effect, photon-matter interaction has
been studied over 100 years. With the establishment of the special
relativity and quantum theory, scientists can make many accurate calculations
to describe how photons are absorbed and emitted and how electrons
are ionized and captured. Most of the work in early years were based
on perturbative techniques, as the light source was so weak that only
single photon effect was important. The accuracy maintained until
the invention of chirped pulse amplification (CPA) for lasers in 1980s.
Since then, the laser power density have increased 8 orders of magnitude,
approaching $10^{22}$ W$\cdot$cm$^{-2}$, which is stronger than
the direct ionization threshold of $10^{16}\sim10^{18}$ W$\cdot$cm$^{-2}$
\cite{Mourou2007,Krausz2009}. Such a strong field brings many new
physics, e.g., multiphoton ionization, above threshold ionization
(ATI), high harmonic generation (HHG) and stabilization, which play
a major role in modern high energy density physics, experimental astrophysics,
attosecond physics, strong field electrodynamics and controlled fusion
etc. \cite{Voronov1965,Keldysh1965,Faisal1973,Agostini1979,Reiss1980,Gontier1980,McPherson1987,Gallagher1988,Eberly1991,Pont1990,Krause1992,Eberly1993,Corkum1993,Lewenstein1994,Birula1994,Bao1996,Spielmann1997,Sali2001,Kienberger2004,Drake2006,Brabec2008,Kim2008,Smirnova2009,Le2009,Goulielmakis2010,Nepstad2010,Birkeland2010,Popmintchev2012,Piazza2012,Becker2012,Madsen2012,Argenti2013,Yuan2013,Guo2013,Klaiber2013,Vampa2014,Popruzhenko2014,Popmintchev2015,Kfir2015,Luu2015,Bukov2015,Hassan2016}.
There are several semi-classical non-perturbative methods to describe
these phenomena, both analytical and numerical, and some experimental
observations have been explained successfully \cite{Keldysh1965,Faisal1973,Reiss1980,Gallagher1988,Krause1992,Corkum1993,Birula1994,Lewenstein1994,Bao1996,Sali2001,Brabec2008,Le2009,Nepstad2010,Becker2012,Klaiber2013}.
Keldysh proposed the first non-perturbative theory describing the
ionization process in a strong laser field \cite{Keldysh1965}. It
was then developed by Faisal and Reiss in the $S$ matrix form known
as KFR theory \cite{Faisal1973,Reiss1980}. This theory was further
developed into the rescattering methods \cite{Bao1996,Le2009}. Simple
man model is a classical model which gives an intuitive perspective
to understand the ionization \cite{Gallagher1988}. In semi-classical
framework, the well known three-step model developed by Corkum gives
a basic tool to study the strong field physics \cite{Corkum1993,Corkum11}.
There are also some models developed based on quantum path-integral
theory which output some detailed results about the transient paths
\cite{Lewenstein1994,Sali2001}. Recently, relativistic corrections
for strong field ionization was taken into consideration by Klaiber
\emph{et al.} \cite{Klaiber2013}. Different from analytical models, directly
solving the time-dependent Schr\"{o}dinger equation (TDSE) is always
a crucially important method for photon-matter interactions. By numerical
simulations, Krause \emph{et al.} obtained the cut-off law of HHG \cite{Krause1992}.
Nepstad \emph{et al.} numerically studied the two-photon ionization of helium
\cite{Nepstad2010}. Birkeland \emph{et al.} numerically studied
the stabilization of helium in intense XUV laser fields \cite{Birkeland2010}.
Based on simulation results, much information about atom and molecular
in strong field can be obtained \cite{Madsen2012,Argenti2013}. Recently,
multi-configuration methods were introduced into TDSE simulations
to treat many-electron dynamics \cite{Hochstuhl2014,Miyagi2014,Bauch2014}.
Because of the multi-scale nature of the process and the large number
of degrees of freedom involved, most of the theoretical and numerical
methods adopted various types of approximations for the Schr\"{o}dinger
equation, such as the strong field approximation \cite{Lewenstein1994},
the finite energy levels approximation \cite{Wu1995}, the independent
external field approximation \cite{Popruzhenko2014} and the single-active
electron approximation \cite{Krause1992}, which often have limited
applicabilities \cite{Krausz2009,Piazza2012}. To understand the
intrinsic multi-scale, complex photon-matter interactions described
by the Schr\"{o}dinger-Maxwell (SM) equations, a comprehensive model
needs to be developed by numerically solving the SM equations.

For the Maxwell equations, many numerical methods, such as the finite-difference
time-domain method has been developed \cite{Yee1966,Mur1981,Berenger1994}.
For the Schr\"{o}dinger equation, unitary algorithm has been proposed
\cite{Wu1995,Blanes2006,Kormann2008,Shen2013}. Recently, a class
of structure-preserving geometric algorithms have been developed for
simulating classical particle-field interactions described by the
Vlasov-Maxwell (VM) equations. Specifically, spatially discretized
canonical and non-canonical Poisson brackets for the VM systems
and associated symplectic time integration algorithms have been discovered
and applied \cite{Squire12,JXiao2013,JXiao2015,Xiao15-112504,He15-124503,QHong2016,He16-092108,Xiao-M2016,Morrison2017,Michael-ar}.

In this paper, we develop a new structure-preserving geometric algorithm
for numerically solving the SM equations. For this purpose, the canonical
symplectic structure of the SM equations is first established. Note
that the canonical symplectic structure presented here is more transparent
than the version given in Refs. \cite{Masiello2004,Masiello2005},
which involves complications due to a different choice of gauge. The
structure-preserving geometric algorithm is obtained by discretizing
the canonical Poisson bracket. The wavefunctions and gauge field are
discretized point-wise on an Eulerian spatial grid, and the Hamiltonian
functional is expressed as a function of the discretized fields. This
procedure generates a finite-dimensional Hamiltonian system with a
canonical symplectic structure. The degrees of freedom of the discrete
system for a single electron atom discrete system is $4M$, where
$M$ is the number of grid points. For an ensemble of $N$ single-active
electron atoms, the discrete system has $(N+3)M$ degrees of freedom.
A symplectic splitting algorithm is developed for semi-explicit time advance.
The method inherits all the good numerical features of canonical symplectic
algorithms, such as the long-term bound on energy-momentum error.
We also design the algorithm such that it preserves unitary structure
of the Schr\"{o}dinger equation. These desirable features make the
algorithm a powerful tool in the study of photon-matter interactions
using the semi-classical model. We note the algorithm developed here
for the SM equations is inspired by the recent advances in the structure-preserving
geometric algorithms for classical particle-field interactions \cite{Squire12,JXiao2013,JXiao2015,Xiao15-112504,He15-124503,QHong2016,He16-092108,Xiao-M2016,Morrison2017,Michael-ar},
especially the canonical particle-in-cell method \cite{QHong2016}.

\section{Canonical Symplectic Structure of Schr\"{o}dinger-Maxwell Systems}

\label{sec:2}

In most strong field experiments, the atomic ensemble is weakly coupled,
which means that electrons are localized around the nuclei and there
is no direct coupling between different atoms. Electrons belong to
different atoms are well resolved. In a single-active electron atomic
ensemble, every electron can be labeled by a local atom potential.
The wavefunction is a direct product of the resolved single electron
wavefunctions. As the basic semi-classical model for photon-matter
interactions between atomic ensemble and photons, the SM equations
are 
\begin{eqnarray}
i\frac{\partial}{\partial{t}}\psi_{i} & = & \hat{H}_{i}\psi_{i},\label{eq:1}\\
\partial_{\mu}F^{\mu\nu} & = & \sum_{i}\frac{4\pi}{c}J_{i}^{\nu},\label{eq:2}
\end{eqnarray}
where $\hat{H}_{i}=\frac{(\bm{P}-\bm{A})^{2}}{2}+V_{i}$ is the Hamiltonian
operator, $\bm{P}=-i\bigtriangledown$ is the canonical momentum,
$V_{i}$ is local atomic potential of the $i$-th atom, $F^{\mu\nu}=c(\partial^{\mu}A^{\nu}-\partial^{\nu}A^{\mu})$
is the electromagnetic tensor, and $c$ is the light speed in atomic
units. The subscript $i$ is electron label. The atomic potential
can assume, for example, the form of $V_{i}(\bm{x})=-\frac{Z}{|\bm{x}-\bm{x}_{i}|}$
with $Z$ being atomic number and $\bm{x}_{i}$ the position of the
atom. With metric signature $\left(+,-,-,-\right)$, in Eq.\,\eqref{eq:2},
$J_{i}^{\mu}=i\left[\psi_{i}^{*}D^{\mu}\psi_{i}-\psi_{i}(D^{\mu}\psi_{i})^{*}\right]$
is the conserved Noether current, and $D_{\mu}=\partial_{\mu}+iA_{\mu}$
is the gauge-covariant derivative. In the nonrelativistic limit, the
density $J_{i}^{0}$ reduces to $\psi_{i}^{*}\psi_{i}$, while the
current density $J_{i}^{k}$ reduces to $\frac{i}{2}\left[\psi_{i}^{*}D^{k}\psi_{i}-\psi_{i}(D^{k}\psi_{i})^{*}\right]$,
which closes the SM system. The temporal gauge
$\phi=0$ has been adopted explicitly.

The complex wavefunctions and Hamiltonian operators can be decomposed
into real and imaginary parts, 
\begin{gather}
\psi_{i}=\frac{1}{\sqrt{2}}\left(\psi_{iR}+i\psi_{iI}\right),\label{eq:3}\\
\hat{H}_{i}=\hat{H}_{iR}+i\hat{H}_{iI},\\
\hat{H}_{iR}=\frac{1}{2}\left(-\bigtriangledown^{2}+\bm{A}^{2}\right)+V_{i},\thinspace\thinspace\thinspace\hat{H}_{iI}=\frac{1}{2}\bigtriangledown\cdot\bm{A}+\bm{A}\cdot\bigtriangledown.
\end{gather}
In terms of the real and imaginary components, the Schr\"{o}dinger
equation is 
\begin{eqnarray}
\frac{\partial}{\partial{t}}\left(\begin{array}{c}
\psi_{iR}\\
\psi_{iI}
\end{array}\right)=\left(\begin{array}{cc}
\hat{H}_{iI} & \hat{H}_{iR}\\
-\hat{H}_{iR} & \hat{H}_{iI}
\end{array}\right)\left(\begin{array}{c}
\psi_{iR}\\
\psi_{iI}
\end{array}\right).\label{eq:7}
\end{eqnarray}

The SM system admits an infinite dimensional canonical symplectic
structure with following Poisson structure and Hamiltonian functional,
\begin{eqnarray}
\left\{ F,G\right\}  & \!=\! & \int\left[\sum_{i}\left(\frac{\delta{F}}{\delta\psi_{iR}}\frac{\delta{G}}{\delta\psi_{iI}}\!-\!\frac{\delta{G}}{\delta\psi_{iR}}\frac{\delta{F}}{\delta\psi_{iI}}\right)\!+\!\frac{\delta{F}}{\delta\bm{A}}\frac{\delta{G}}{\delta\bm{Y}}\!-\!\frac{\delta{G}}{\delta\bm{A}}\frac{\delta{F}}{\delta\bm{Y}}\right]\mathrm{d}^{3}x,\label{eq:10}\\
H\left(\psi_{iR},\psi_{iI},\bm{A},\bm{Y}\right) & \!=\! & \frac{1}{2}\int\left[\sum_{i}\left(\psi_{iR}\hat{H}_{iR}\psi_{iR}\!+\!\psi_{iI}\hat{H}_{iR}\psi_{iI}\right.\right.\nonumber \\
 & \, & \left.\left.\!-\!\psi_{iR}\hat{H}_{iI}\psi_{iI}\!+\!\psi_{iI}\hat{H}_{iI}\psi_{iR}\right)\!+\!4\pi\bm{Y}^{2}\!+\!\frac{1}{4\pi}\left(c\bigtriangledown\!\times\!\bm{A}\right)^{2}\right]\mathrm{d}^{3}x.\label{eq:11}
\end{eqnarray}
Here, $\bm{Y}=\dot{\bm{A}}/4\pi$ and $F,$ $G,$ and $H$ are functionals
of $\left(\psi_{iR},\psi_{iI},\bm{A},\bm{Y}\right).$ The expression
$\delta F/\delta\psi_{iR}$ is the variational derivative of the functional
$F$ with respect to $\psi_{iR}$, and other terms, e.g., $\delta F/\delta\psi_{iI}$
and $\delta F/\delta\boldsymbol{A}$, have similar meanings. The Hamiltonian
functional $H\left(\psi_{iR},\psi_{iI},\bm{A},\bm{Y}\right)$ in Eq.\,\eqref{eq:11}
is equivalent to the following expression in terms of the complex
wavefunctions, 
\begin{eqnarray}
H\left(\psi_{i}^{*},\psi_{i},\bm{A},\bm{Y}\right) & = & H_{qm}+H_{em},\label{eq:8}\\
H_{qm} & = & \int\sum_{i}\psi_{i}^{*}\hat{H}_{i}\psi_{i}\mathrm{d}^{3}x,\\
H_{em} & = & \frac{1}{2}\int\left[4\pi\bm{Y}^{2}+\frac{1}{4\pi}\left(c\bigtriangledown\times\bm{A}\right)^{2}\right]\mathrm{d}^{3}x.
\end{eqnarray}
Apparently, $H_{em}$ is the Hamiltonian for the electromagnetic field,
and $H_{qm}$ is the Hamiltonian for the wavefunctions. In this infinite
dimensional Hamiltonian system, the canonical pairs are $\left(\psi_{iR},\psi_{iI}\right)$
and $\left(\bm{A},\bm{Y}\right)$ at each spatial location. Their
canonical equations are 
\begin{eqnarray}
\dot{\psi_{iR}} & = & \left\{ \psi_{iR},H\right\} =\!\frac{1}{2}\bigtriangledown\cdot\bm{A}\psi_{iR}\!+\!\bm{A}\cdot\bigtriangledown\psi_{iR}\!+\!\frac{1}{2}\left(\!-\!\bigtriangledown^{2}\!+\!\bm{A}^{2}\right)\psi_{iI}\!+\!V_{i}\psi_{iI}\!,\label{eq:13}\\
\dot{\bm{A}} & = & \left\{ \bm{A},H\right\} =4\pi\bm{Y},\label{eq:14}\\
\dot{\psi_{iI}} & = & \left\{ \psi_{iI},H\right\} =\frac{1}{2}\left(\bigtriangledown^{2}\!-\!\bm{A}^{2}\right)\psi_{iR}\!-\!V_{i}\psi_{iR}\!+\!\frac{1}{2}\bigtriangledown\cdot\bm{A}\psi_{iI}\!+\!\bm{A}\cdot\bigtriangledown\psi_{iI},\label{eq:15}\\
\dot{\bm{Y}} & = & \left\{ \bm{Y},H\right\} =\bm{\mathcal{J}}\!-\!\frac{c^{2}}{4\pi}\bigtriangledown\times\bigtriangledown\times\bm{A},\label{eq:16}
\end{eqnarray}
where $\bm{\mathcal{J}}=\frac{1}{2}\sum_{i}[\psi_{iR}\bigtriangledown\psi_{iI}-\psi_{iI}\bigtriangledown\psi_{iR}-(\psi_{iR}^{2}+\psi_{iI}^{2})\bm{A}]$
is the current density. In deriving Eqs.\,\eqref{eq:13}-\eqref{eq:16},
use is made of the following expression of the total variation of
Hamiltonian, 
\begin{eqnarray}
\delta H & = & \frac{1}{2}\int\sum_{i}[\left(-\bigtriangledown^{2}\psi_{iR}+\bm{A}^{2}\psi_{iR}+2V_{i}\psi_{iR}-2\bm{A}\cdot\bigtriangledown\psi_{iI}-\bigtriangledown\cdot\bm{A}\psi_{iI}\right)\delta\psi_{iR}\nonumber \\
 & \, & +\left(-\bigtriangledown^{2}\psi_{iI}+\bm{A}^{2}\psi_{iI}+2V_{i}\psi_{iI}+2\bm{A}\cdot\bigtriangledown\psi_{iR}+\bigtriangledown\cdot\bm{A}\psi_{iR}\right)\delta\psi_{iI}\nonumber \\
 & \, & +\left(\psi_{iR}^{2}\bm{A}+\psi_{iI}^{2}\bm{A}+\psi_{iI}\bigtriangledown\psi_{iR}-\psi_{iR}\bigtriangledown\psi_{iI}\right)\cdot\delta\bm{A}]\mathrm{d}^{3}x\nonumber \\
 & \, & +\int[\frac{c^{2}}{4\pi}\bigtriangledown\times\bigtriangledown\times\bm{A}\cdot\delta\bm{A}+4\pi\bm{Y}\cdot\delta\bm{Y}]\mathrm{d}^{3}x,\label{eq:12}
\end{eqnarray}
where integration by parts have been applied with fixed fields on
the boundary.

\section{Structure-preserving Geometric Algorithms for Schr\"{o}dinger-Maxwell
Systems}

\label{sec:3}

We now present the structure-preserving geometric algorithms for numerically
solving Eqs.\,\eqref{eq:13}-\eqref{eq:16}. We discretize the fields
$\left(\psi_{iR},\psi_{iI},\bm{A},\bm{Y}\right)$ on an Eulerian spatial
grid as 
\begin{eqnarray}
\bm{A}\left(\bm{x},t\right)=\sum_{J=1}^{M}\bm{A}_{J}\left(t\right)\theta\left(\bm{x}-\bm{x}_{J}\right) & ,\,\,\, & \bm{Y}\left(\bm{x},t\right)=\sum_{J=1}^{M}\bm{Y}_{J}\left(t\right)\theta\left(\bm{x}-\bm{x}_{J}\right),\label{eq:17}\\
\psi_{iR}\left(\bm{x},t\right)=\sum_{J=1}^{M}\psi_{iRJ}\left(t\right)\theta\left(\bm{x}-\bm{x}_{J}\right) & ,\,\,\, & \psi_{iI}\left(\bm{x},t\right)=\sum_{J=1}^{M}\psi_{iIJ}\left(t\right)\theta\left(\bm{x}-\bm{x}_{J}\right),\label{eq:18}
\end{eqnarray}
where the distribution function $\theta\left(\bm{x}-\bm{x}_{J}\right)$
is defined as 
\begin{eqnarray}
\theta\left(\bm{x}-\bm{x}_{J}\right)=\left\{ \begin{array}{cc}
1, & |x-x_{J}|<\frac{\bigtriangleup{x}}{2},|y-y_{J}|<\frac{\bigtriangleup{y}}{2},|z-z_{J}|<\frac{\bigtriangleup{z}}{2}\\
0, & elsewhere
\end{array}\right..\label{eq:19}
\end{eqnarray}
Then, the variational derivative with respect to $\bm{A}$ is 
\begin{eqnarray}
\frac{\delta{F}}{\delta\bm{A}}=\sum_{J=1}^{M}\frac{\delta\bm{A}_{J}}{\delta\bm{A}}\frac{\partial{F}}{\partial\bm{A}_{J}}=\sum_{J=1}^{M}\frac{1}{\bigtriangleup{V}}\theta\left(\bm{x}-\bm{x}_{J}\right)\frac{\partial{F}}{\partial\bm{A}_{J}},\label{eq:20}
\end{eqnarray}
and the variational derivatives with respect to $\bm{Y}$, $\psi_{iR}$
and $\psi_{iI}$ have similar expressions. Here, $\bigtriangleup V=\bigtriangleup x\bigtriangleup y\bigtriangleup z$
is the volume of each cell. The canonical Poisson bracket is discretized
as 
\begin{eqnarray}
\left\{ F,G\right\} _{d}\!=\!\sum_{J=1}^{M}\left[\sum_{i}\left(\frac{\partial{F}}{\partial\psi_{iRJ}}\frac{\partial{G}}{\partial\psi_{iIJ}}\!-\!\frac{\partial{G}}{\partial\psi_{iRJ}}\frac{\partial{F}}{\partial\psi_{iIJ}}\right)\!+\!\frac{\partial{F}}{\partial\bm{A}_{J}}\frac{\partial{G}}{\partial\bm{Y}_{J}}\!-\!\frac{\partial{G}}{\partial\bm{A}_{J}}\frac{\partial{F}}{\partial\bm{Y}_{J}}\right]\frac{1}{\bigtriangleup{V}}.\label{eq:21}
\end{eqnarray}
The Hamiltonian functional is discretized as 
\begin{gather}
H_{d}\left(\psi_{iRJ},\psi_{iIJ},\bm{A}_{J},\bm{Y}_{J}\right)=H_{dem}+H_{dqm},\label{eq:Hds-1}\\
H_{dem}=\frac{1}{2}\sum_{J=1}^{M}\left[4\pi\bm{Y}_{J}^{2}\!+\!\frac{1}{4\pi}\left(c\bigtriangledown_{d}\!\times\!\bm{A}\right)_{J}^{2}\right]\bigtriangleup V,\label{eq:Hdem-1}\\
H_{dqm}=\frac{1}{2}\sum_{J=1}^{M}\sum_{i}\left[-\frac{1}{2}\psi_{iRJ}\left(\bigtriangledown_{d}^{2}\psi_{iR}\right)_{J}\!-\!\frac{1}{2}\psi_{iIJ}\left(\bigtriangledown_{d}^{2}\psi_{iI}\right)_{J}\!-\!\psi_{iRJ}\bm{A}_{J}\cdot\left(\bigtriangledown_{d}\psi_{iI}\right)_{J}\right.\nonumber \\
\left.+\psi_{iIJ}\bm{A}_{J}\cdot\left(\bigtriangledown_{d}\psi_{iR}\right)_{J}\!+\!\left(\frac{1}{2}\bm{A}_{J}^{2}\!+\!V_{iJ}\right)\left(\psi_{iRJ}^{2}\!+\!\psi_{iIJ}^{2}\right)\right]\bigtriangleup V,\label{eq:Hdqm-1}
\end{gather}
where $V_{iJ}=V_{i}\left(\bm{x}_{J}\right)$, and the discrete spatial
operators are defined as 
\begin{eqnarray}
\left(\bigtriangledown_{d}\psi\right)_{J}=\left(\begin{array}{c}
\frac{\psi_{i,j,k}-\psi_{i-1,j,k}}{\bigtriangleup{x}}\\
\frac{\psi_{i,j,k}-\psi_{i,j-1,k}}{\bigtriangleup{y}}\\
\frac{\psi_{i,j,k}-\psi_{i,j,k-1}}{\bigtriangleup{z}}
\end{array}\right),\label{eq:23}
\end{eqnarray}
\begin{eqnarray}
\left(\bigtriangledown_{d}\cdot\bm{A}\right)_{J}=\frac{Ax_{i,j,k}-Ax_{i-1,j,k}}{\bigtriangleup{x}}+\frac{Ay_{i,j,k}-Ay_{i,j-1,k}}{\bigtriangleup{y}}+\frac{Az_{i,j,k}-Az_{i,j,k-1}}{\bigtriangleup{z}},\label{eq:24}
\end{eqnarray}
\begin{eqnarray}
\left(\bigtriangledown_{d}\times\bm{A}\right)_{J}=\left(\begin{array}{c}
\frac{Az_{i,j,k}-Az_{i,j-1,k}}{\bigtriangleup{y}}-\frac{Ay_{i,j,k}-Ay_{i,j,k-1}}{\bigtriangleup{z}}\\
\frac{Ax_{i,j,k}-Ax_{i,j,k-1}}{\bigtriangleup{z}}-\frac{Az_{i,j,k}-Az_{i-1,j,k}}{\bigtriangleup{x}}\\
\frac{Ay_{i,j,k}-Ay_{i-1,j,k}}{\bigtriangleup{x}}-\frac{Ax_{i,j,k}-Ax_{i,j-1,k}}{\bigtriangleup{y}}
\end{array}\right),\label{eq:25}
\end{eqnarray}
\begin{eqnarray}
\left(\bigtriangledown_{d}^{2}\psi\right)_{J} & = & \frac{\psi_{i,j,k}\!-\!2\psi_{i-1,j,k}\!+\!\psi_{i-2,j,k}}{\bigtriangleup{x}^{2}}\!+\!\frac{\psi_{i,j,k}\!-\!2\psi_{i,j-1,k}\!+\!\psi_{i,j-2,k}}{\bigtriangleup{y}^{2}}\nonumber \\
 &  & +\frac{\psi_{i,j,k}\!-\!2\psi_{i,j,k-1}\!+\!\psi_{i,j,k-2}}{\bigtriangleup{z}^{2}}.\label{eq:26}
\end{eqnarray}
Here, the subscript $J$ denotes grid position $(i,j,k)$. The discrete
spatial operators defined here use first order backward difference
schemes. High order spatial schemes can be developed as well.

The discrete canonical equations are 
\begin{eqnarray}
\dot{\psi_{iRJ}} & = & \left\{ \psi_{iRJ},H_{d}\right\} _{d}\nonumber \\
 & = & \frac{1}{2}\bm{A}_{J}\cdot\left(\bigtriangledown_{d}\psi_{iR}\right)_{J}\!-\!\frac{1}{2}\sum_{K=1}^{M}\psi_{iRK}\bm{A}_{K}\cdot\frac{\partial}{\partial\psi_{iIJ}}\left(\bigtriangledown_{d}\psi_{iI}\right)_{K}\nonumber \\
 &  & -\frac{1}{4}\left(\bigtriangledown_{d}^{2}\psi_{iI}\right)_{J}\!-\!\frac{1}{4}\sum_{K=1}^{M}\psi_{iIK}\frac{\partial}{\partial\psi_{iIJ}}\left(\bigtriangledown_{d}^{2}\psi_{iI}\right)_{K}\!+\!\left(\frac{1}{2}\bm{A}_{J}^{2}\!+\!V_{iJ}\right)\psi_{iIJ},\label{eq:27}\\
\dot{\bm{A}_{J}} & = & \left\{ \bm{A}_{J},H_{d}\right\} _{d}=4\pi\bm{Y}_{J},\label{eq:28}\\
\dot{\psi_{iIJ}} & = & \left\{ \psi_{iIJ},H_{d}\right\} _{d}\nonumber \\
 & = & \frac{1}{4}\left(\bigtriangledown_{d}^{2}\psi_{iR}\right)_{J}\!+\!\frac{1}{4}\sum_{K=1}^{M}\psi_{iRK}\frac{\partial}{\partial\psi_{iRJ}}\left(\bigtriangledown_{d}^{2}\psi_{iR}\right)_{K}\!-\!\left(\frac{1}{2}\bm{A}_{J}^{2}\!+\!V_{iJ}\right)\psi_{iRJ}\nonumber \\
 &  & +\frac{1}{2}\bm{A}_{J}\cdot\left(\bigtriangledown_{d}\psi_{iI}\right)_{J}\!-\!\frac{1}{2}\sum_{K=1}^{M}\psi_{iIK}\bm{A}_{K}\cdot\frac{\partial}{\partial\psi_{iRJ}}\left(\bigtriangledown_{d}\psi_{iR}\right)_{K},\label{eq:29}\\
\dot{\bm{Y}_{J}} & = & \left\{ \bm{Y}_{J},H_{d}\right\} _{d}=\bm{\mathcal{J}}_{J}\!-\!\frac{c^{2}}{4\pi}\left(\bigtriangledown_{d}^{T}\times\bigtriangledown_{d}\times\bm{A}\right)_{J},\label{eq:30}
\end{eqnarray}
where $\bm{\mathcal{J}}_{J}=\frac{1}{2}\sum_{i}[\psi_{iRJ}\left(\bigtriangledown_{d}\psi_{iI}\right)_{J}-\psi_{iIJ}\left(\bigtriangledown_{d}\psi_{iR}\right)_{J}-\bm{A}_{J}\left(\psi_{iRJ}^{2}+\psi_{iIJ}^{2}\right)]$
is the discrete current density. The last term in Eq.~\eqref{eq:30}
is defined to be, 
\begin{eqnarray}
\left(\bigtriangledown_{d}^{T}\times\bigtriangledown_{d}\times\bm{A}\right)_{J}=\frac{1}{2}\frac{\partial}{\partial\bm{A}_{J}}\left[\sum_{K=1}^{M}\left(\bigtriangledown_{d}\times\bm{A}\right)_{K}^{2}\right],\label{eq:yinyong}
\end{eqnarray}
which indicates that the right-hand side of Eq.~\eqref{eq:yinyong}
can be viewed as the discretized $\bigtriangledown\times\bigtriangledown\times\bm{A}$
for a well-chosen discrete curl operator $\bigtriangledown_{d}\times$.

We will use the following symplectic splitting algorithms to numerically
solve this set of discrete canonical Hamiltonian equations. In Eq.\,\eqref{eq:Hds-1},
$H_{d}$ is naturally split into two parts, each of which corresponds
to a subsystem that will be solved independently. The solution maps
of the subsystems will be combined in various way to give the desired
algorithms for the full system. For the subsystem determined by $H_{dqm},$
the dynamic equations are 
\begin{eqnarray}
\dot{\psi_{iRJ}} & = & \left\{ \psi_{iRJ},H_{dqm}\right\} _{d}\nonumber \\
 & = & \frac{1}{2}\bm{A}_{J}\cdot\left(\bigtriangledown_{d}\psi_{iR}\right)_{J}\!-\!\frac{1}{2}\sum_{K=1}^{M}\psi_{iRK}\bm{A}_{K}\cdot\frac{\partial}{\partial\psi_{iIJ}}\left(\bigtriangledown_{d}\psi_{iI}\right)_{K}\nonumber \\
 &  & -\frac{1}{4}\left(\bigtriangledown_{d}^{2}\psi_{iI}\right)_{J}\!-\!\frac{1}{4}\sum_{K=1}^{M}\psi_{iIK}\frac{\partial}{\partial\psi_{iIJ}}\left(\bigtriangledown_{d}^{2}\psi_{iI}\right)_{K}\!+\!\left(\frac{1}{2}\bm{A}_{J}^{2}\!+\!V_{iJ}\right)\psi_{iIJ},\label{eq:psiRJ}\\
\dot{\psi_{iIJ}} & = & \left\{ \psi_{iIJ},H_{dqm}\right\} _{d}\nonumber \\
 & = & \frac{1}{4}\left(\bigtriangledown_{d}^{2}\psi_{iR}\right)_{J}\!+\!\frac{1}{4}\sum_{K=1}^{M}\psi_{iRK}\frac{\partial}{\partial\psi_{iRJ}}\left(\bigtriangledown_{d}^{2}\psi_{iR}\right)_{K}\!-\!\left(\frac{1}{2}\bm{A}_{J}^{2}\!+\!V_{iJ}\right)\psi_{iRJ}\nonumber \\
 &  & +\frac{1}{2}\bm{A}_{J}\cdot\left(\bigtriangledown_{d}\psi_{iI}\right)_{J}\!-\!\frac{1}{2}\sum_{K=1}^{M}\psi_{iIK}\bm{A}_{K}\cdot\frac{\partial}{\partial\psi_{iRJ}}\left(\bigtriangledown_{d}\psi_{iR}\right)_{K},\label{eq:psiIJ}\\
\dot{\bm{A}_{J}} & = & \left\{ \bm{A}_{J},H_{dqm}\right\} _{d}=0,\\
\dot{\bm{Y}_{J}} & = & \left\{ \bm{Y}_{J},H_{dqm}\right\} _{d}=\bm{\mathcal{J}}_{J}.\!\label{eq:30-1}
\end{eqnarray}
Equations \eqref{eq:psiRJ} and \eqref{eq:psiIJ} can written as 
\begin{gather}
\frac{d}{dt}\left(\begin{array}{c}
\psi_{iR}\\
\psi_{iI}
\end{array}\right)=\Omega(\boldsymbol{A})\left(\begin{array}{c}
\psi_{iR}\\
\psi_{iI}
\end{array}\right),
\end{gather}
where $\Omega(\boldsymbol{A})$ is an skew-symmetric matrix. It easy
to show that $\Omega(\boldsymbol{A})$ is also an infinitesimal generator
of the symplectic group. To preserve the unitary property of $\psi_{i}$,
we adopt the symplectic mid-point method for this subsystem, and the
one step map $M_{qm}:(\psi_{i},\boldsymbol{A},\boldsymbol{Y})^{n}\longmapsto(\psi_{i},\boldsymbol{A},\boldsymbol{Y})^{n+1}$
is given by 
\begin{gather}
\left(\begin{array}{c}
\psi_{iR}\\
\psi_{iI}
\end{array}\right)^{n+1}=\left(\begin{array}{c}
\psi_{iR}\\
\psi_{iI}
\end{array}\right)^{n}+\frac{\Delta t}{2}\Omega(\boldsymbol{A}^{n})\left[\left(\begin{array}{c}
\psi_{iR}\\
\psi_{iI}
\end{array}\right)^{n}+\left(\begin{array}{c}
\psi_{iR}\\
\psi_{iI}
\end{array}\right)^{n+1}\right],\label{eq:psin+1}\\
\boldsymbol{A}^{n+1}=\boldsymbol{A}^{n},\\
\boldsymbol{Y}^{n+1}=\boldsymbol{Y}^{n}+\Delta{t}\bm{\mathcal{J}}\left(\frac{\psi_{iR}^{n}+\psi_{iR}^{n+1}}{2},\frac{\psi_{iI}^{n}+\psi_{iI}^{n+1}}{2}\right).
\end{gather}
Equation \eqref{eq:psin+1} is a linear equation in terms of $(\psi_{iR}^{n+1},\psi_{iI}^{n+1})$.
Its solution is 
\begin{gather}
\left(\begin{array}{c}
\psi_{iR}\\
\psi_{iI}
\end{array}\right)^{n+1}=Cay(\Omega(\boldsymbol{A}^{n})\frac{\Delta t}{2})\left(\begin{array}{c}
\psi_{iR}\\
\psi_{iI}
\end{array}\right)^{n},\\
Cay(\Omega(\boldsymbol{A}^{n})\frac{\Delta t}{2})=\left(1-\Omega(\boldsymbol{A}^{n})\frac{\Delta t}{2}\right)^{-1}\left(1+\Omega(\boldsymbol{A}^{n})\frac{\Delta t}{2}\right),
\end{gather}
where $Cay(S)$ denotes the Cayley transformation of matrix $S.$
It is well-known that $Cay(S)$ is a symplectic rotation matrix when
$S$ in the Lie algebra of the symplectic rotation group. Thus, the
one-step map from $\psi_{i}^{n}=\psi_{iR}^{n}+i\psi_{iI}^{n}$ to
$\psi_{i}^{n+1}=\psi_{iR}^{n+1}+i\psi_{iI}^{n+1}$ induced by $M_{qm}$
for the subsystem $H_{dqm}$ is unitary. Since $\Omega(\boldsymbol{A}^{n}\Delta t/2)$
is a sparse matrix, there exist efficient algorithms to solve Eq.\,\eqref{eq:psin+1}
or to calculate $Cay(\Omega(\boldsymbol{A}^{n})\Delta t/2)$. Once
$\psi_{i}^{n+1}$is known, $\boldsymbol{Y}^{n+1}$ can be calculated
explicitly. Thus, $M_{qm}:(\psi_{i},\boldsymbol{A},\boldsymbol{Y})^{n}\longmapsto(\psi_{i},\boldsymbol{A},\boldsymbol{Y})^{n+1}$
is a second-order symplectic method, which also preserves the unitariness
of $\psi_{i}$.

For the subsystem $H_{dem},$ the dynamic equations are 
\begin{eqnarray}
\dot{\psi_{iRJ}} & = & \left\{ \psi_{iRJ},H_{dem}\right\} _{d}=0,\label{eq:27-1-1}\\
\dot{\psi_{iIJ}} & = & \left\{ \psi_{iIJ},H_{dem}\right\} _{d}=0,\label{eq:29-1-1}\\
\dot{\bm{A}_{J}} & = & \left\{ \bm{A}_{J},H_{dem}\right\} _{d}=4\pi\bm{Y}_{J},\label{eq:AJ}\\
\dot{\bm{Y}_{J}} & = & \left\{ \bm{Y}_{J},H_{dem}\right\} _{d}=-\frac{c^{2}}{4\pi}\left(\bigtriangledown_{d}^{T}\times\bigtriangledown_{d}\times\bm{A}\right)_{J}.\!\label{eq:YJ}
\end{eqnarray}
Equations \eqref{eq:AJ} and \eqref{eq:YJ} are linear in terms of
$\boldsymbol{A}$ and $\boldsymbol{Y}$, and can be written as 
\begin{gather}
\frac{d}{dt}\left(\begin{array}{c}
\boldsymbol{A}\\
\boldsymbol{Y}
\end{array}\right)=Q\left(\begin{array}{c}
\boldsymbol{A}\\
\boldsymbol{Y}
\end{array}\right),
\end{gather}
where $Q$ is a constant matrix. We also use the second order symplectic
mid-point rule for this subsystem, and the one step map $M_{em}:(\psi_{i},\boldsymbol{A},\boldsymbol{Y})^{n}\longmapsto(\psi_{i},\boldsymbol{A},\boldsymbol{Y})^{n+1}$
is given explicitly by 
\begin{gather}
\left(\begin{array}{c}
\psi_{iR}\\
\psi_{iI}
\end{array}\right)^{n+1}=\left(\begin{array}{c}
\psi_{iR}\\
\psi_{iI}
\end{array}\right)^{n},\\
\left(\begin{array}{c}
\boldsymbol{A}\\
\boldsymbol{Y}
\end{array}\right)^{n+1}=Cay\left(Q\frac{\Delta t}{2}\right)\left(\begin{array}{c}
\boldsymbol{A}\\
\boldsymbol{Y}
\end{array}\right)^{n}.
\end{gather}
Since the map does not change $\psi_{i}$, it is unitary.

Given the second-order symmetric symplectic one-step maps $M_{em}$ and $M_{qm}$
for the subsystems $H_{dem}$ and $H_{dqm}$, respectively, various
symplectic algorithms for the system can be constructed by composition.
For example, a first-order algorithm for $H_{d}$ is 
\begin{equation}
M(\Delta t)=M_{em}(\Delta t)\circ M_{qm}(\Delta t).
\end{equation}
A second-order symplectic symmetric method can be constructed by the
following symmetric composition, 
\begin{equation}
M^{2}(\Delta t)=M_{em}(\Delta t/2)\circ M_{qm}(\Delta t)\circ M_{em}(\Delta t/2).
\end{equation}
From a $2l$-th order symplectic symmetric method $M^{2l}(\Delta t)$,
a $2(l+1)$-th order symplectic symmetric method can be constructed
as 
\begin{gather}
M^{2(l+1)}(\Delta t)=M^{2l}(\alpha_{l}\Delta t)\circ M^{2l}(\beta_{l}\Delta t)\circ M^{2l}(\alpha_{l}\Delta t)\thinspace,\\
\text{with~}\alpha_{l}=\left(2-2^{1/(2l+1)}\right)^{-1},\thinspace\textrm{ and }\beta_{l}=1-2\alpha_{l}\thinspace.
\end{gather}
Obviously, the composed algorithms for the full system is symplectic
and unitary.

\section{Numerical Examples}

\label{sec:5} As numerical examples, two semi-classical problems
have been solved using an implementation of the first-order structure-preserving
geometric algorithm described above. Simulations are carried out on
a Scientific Linux 6.3 OS with two 2.1 GHz Intel Core2 CPUs. The data
structure is designed in coordinate sparse format and the BICGSTAB
method (iteration accuracy $10^{-9}$) is introduced to implement
the Cayley transformation. 

\begin{figure}
\includegraphics[width=15cm,height=8cm]{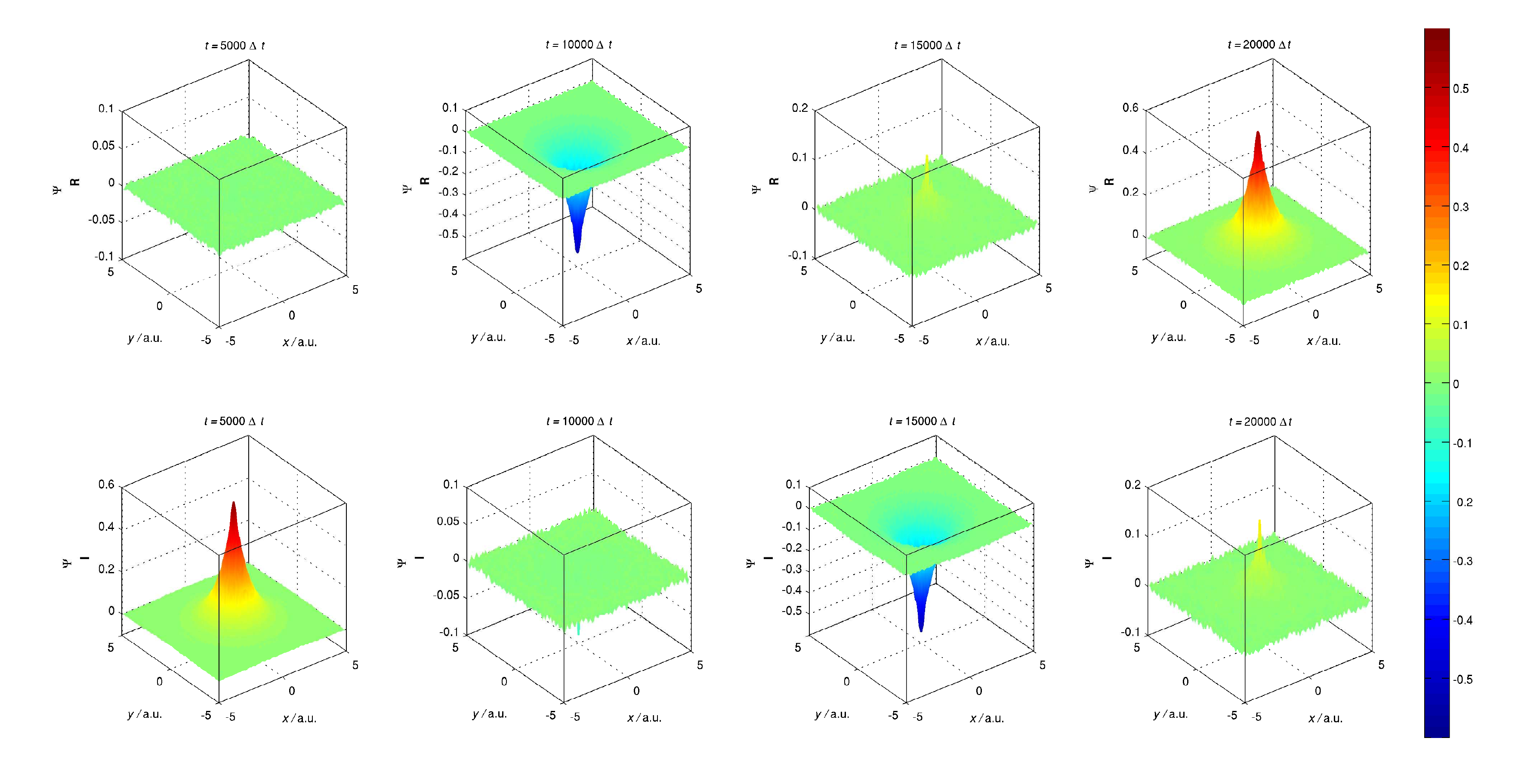} \caption{\label{fig:1} Oscillation of the wavefunction for a hydrogen atom.
The real and imaginary parts of wave function on the $z=0$
plane which passes through the nuclear center. One oscillation cycle
is shown.}
\end{figure}

\begin{figure}
\includegraphics[width=15cm,height=8cm]{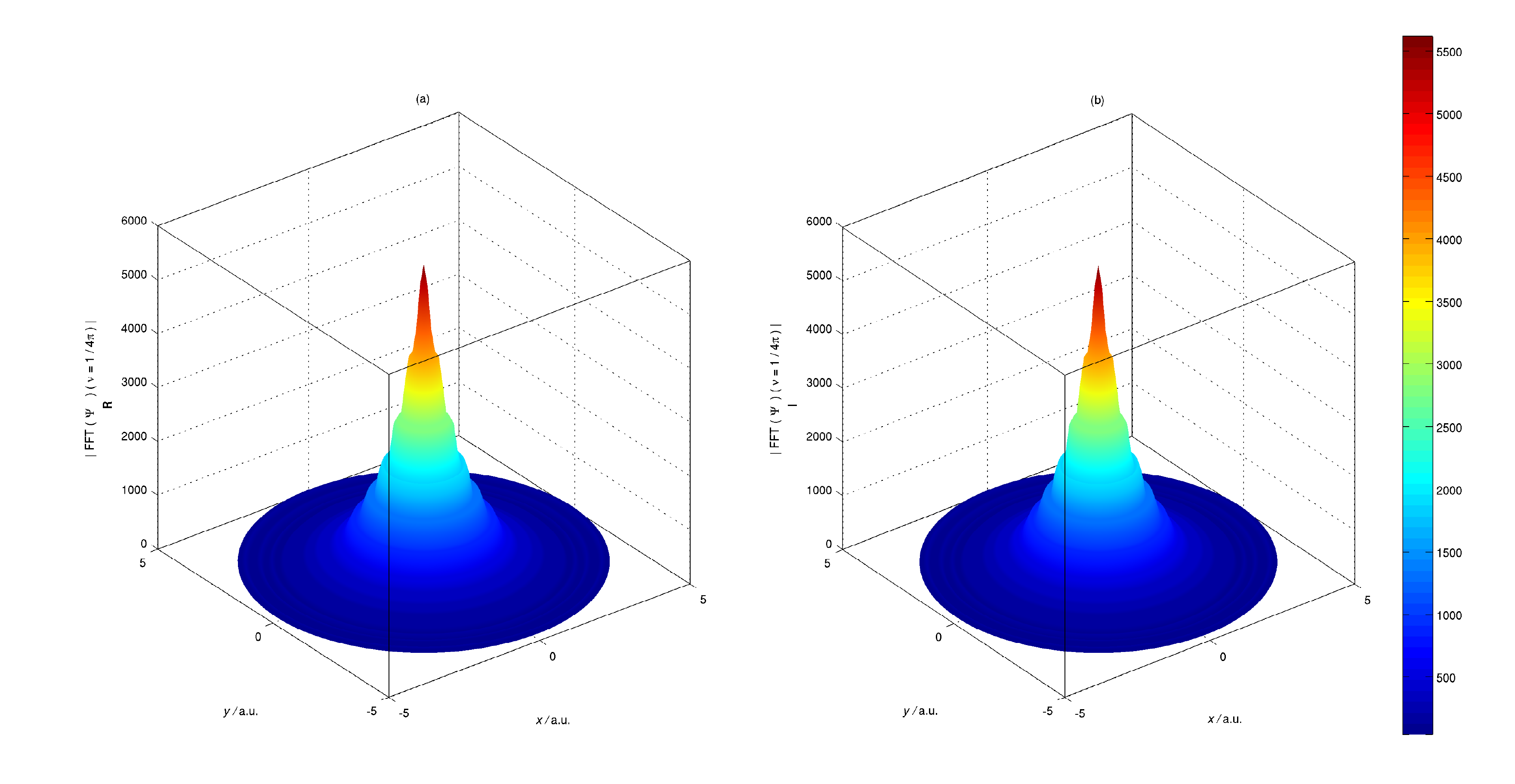} \caption{\label{fig:2} Mode structure of the ground state. Real part (a) and
imaginary part (b) on $z=0$ plane are plotted for the frequency
component at $\nu=1/4\pi$ a.u..}
\end{figure}

\begin{figure}
\includegraphics[width=15cm,height=8cm]{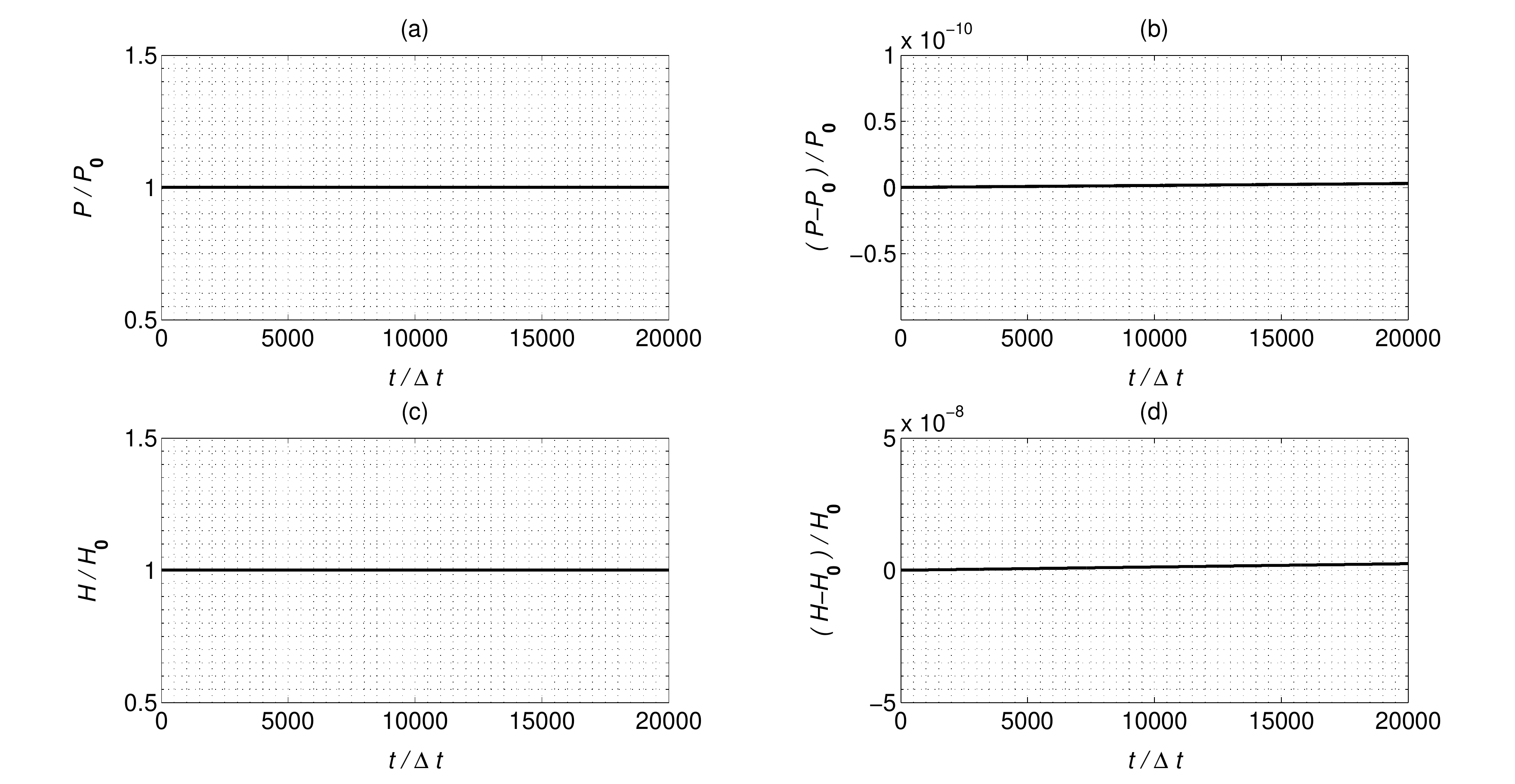} \caption{\label{fig:3} Time-history of numerical errors. After a long-term
simulation, both total probability error (a)-(b) and total Hamiltonian
error (c)-(d) are bounded by a small value.}
\end{figure}

The first numerical example is the oscillation of a free hydrogen
atom, which has been well studied both theoretically and experimentally
\cite{Landau1965,Dirac1958}. The simulation domain is a 100$\times$100$\times$100
uniform Cartesian grid, which represents a {[}-5, 5{]}$\times${[}-5,
5{]}$\times${[}-5, 5{]} $\textrm{a.u.}^{3}$ physical space. All
boundaries are periodic. A hydrogen nucleus is fixed on the origin
and the initial wave function is a direct discretization of the ground-state
wavefunction $\psi=\frac{1}{\sqrt{\pi}}e^{-r}$. The time step is
$\bigtriangleup{t}=1.5\delta/\sqrt{3}c$ a.u., where $\delta=\bigtriangleup{x}=\bigtriangleup{y}=\bigtriangleup{z}=0.1$
a.u. and $c\approx137$ a.u. A total of $2\times10^{4}$ simulation
steps covers a complete oscillation cycle of the ground state. Simulation
results show the ground-state oscillation with very small numerical
noise. Due to the finite-grid size effect and self-field effect, the
initial wave function is not the exact numerical ground state of discrete
hydrogen atom. It is only a good approximation, which couples weakly
to other energy levels. The real and imaginary parts of the wavefunction
on the $z=0$ plane at four different times are plotted in Fig.~\ref{fig:1}.
The numerical oscillation period is found to be 12.58 a.u., which
agrees the analytical result $4\pi$ a.u. very well.
The mode structures at the frequency $\nu=1/4\pi$ a.u. are plotted
in Fig.~\ref{fig:2}. 
As expected, the structure-preserving geometric algorithm has excellent
long-term properties. The time-history of numerical errors are plotted
in Fig.~\ref{fig:3}. After a long-term simulation, both total probability
error and total Hamiltonian error are bounded by a small value.

\begin{figure}
\includegraphics[width=15cm,height=8cm]{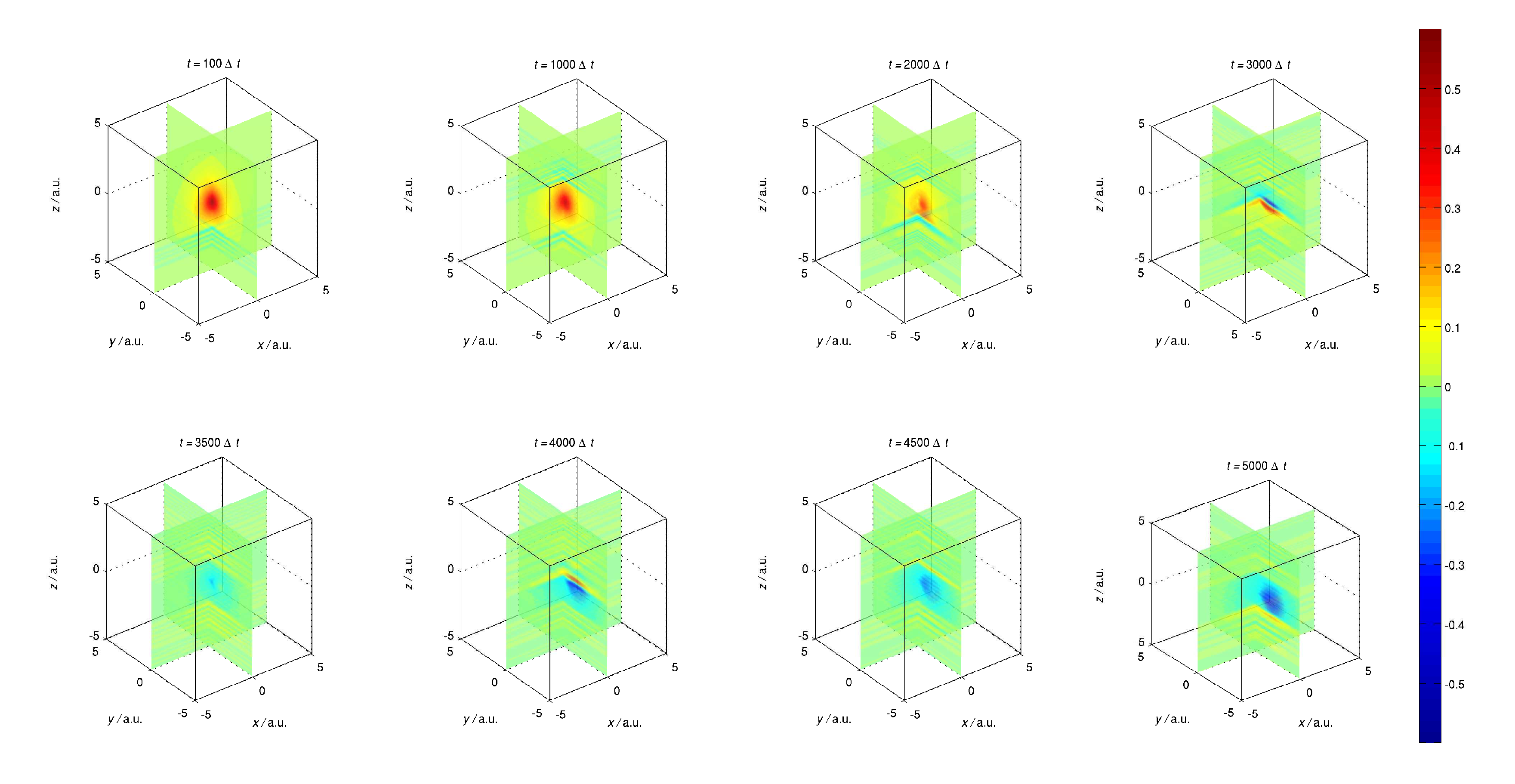} \caption{\label{fig:4} Evolution of the wavefunction (real part). It shows
that at early time, the wave function is localized and the atomic
state is maintained. After a few pulses, the wave function is slightly
modified by the gauge field and plane wave components along the $z$-direction
can be found, which marks the beginning of ionization. With the accumulation
of pulse-train, the wave function drifts along the $\bm{A}\times\bm{k}$
direction, and the atomic state is broken. The increasing plane-wave
components due to ionization can be clearly identified. In this process,
photon momentum is transferred to the electron gradually.}
\end{figure}

\begin{figure}
\includegraphics[width=15cm,height=8cm]{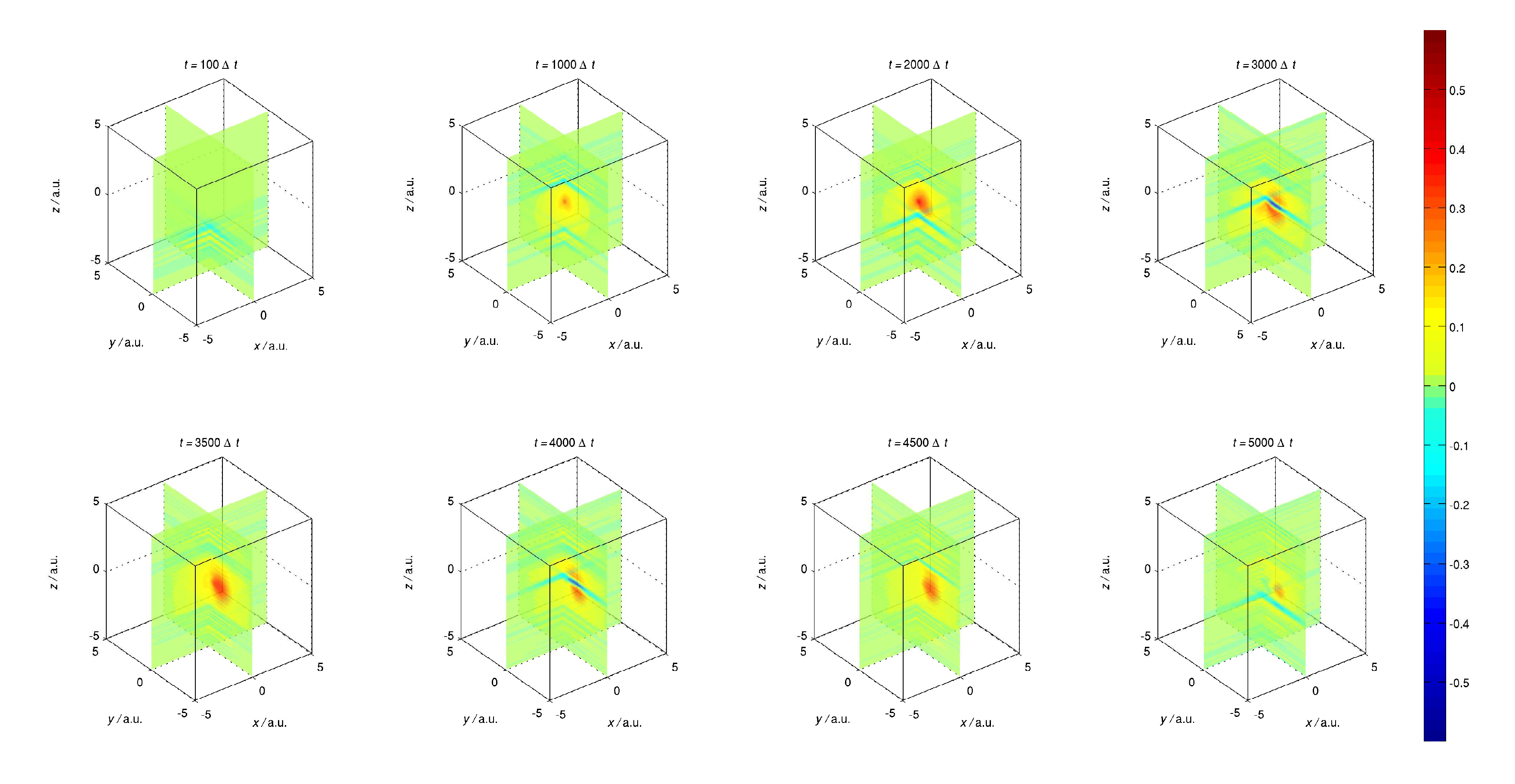} \caption{\label{fig:5} Evolution of the wavefunction (imaginary part). It
shows the same ionization process as in Fig.\,\ref{fig:4}.}
\end{figure}

\begin{figure}
\includegraphics[width=15cm,height=8cm]{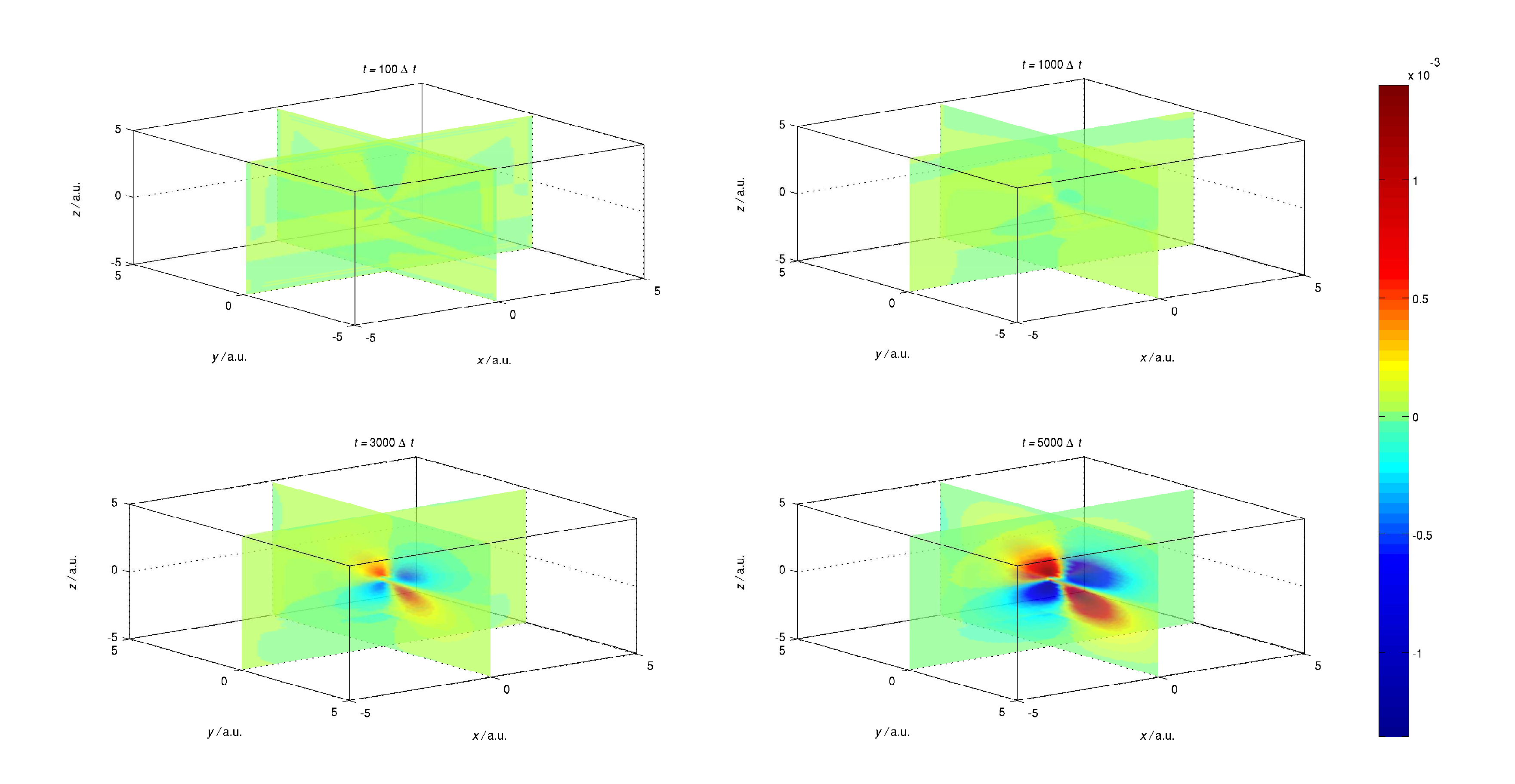} \caption{\label{fig:6} Evolution of the $A_{z}$ component of scattered gauge
field. The scattered field dependents strongly on the electron polarization
current. It is weak relative to the incident field, which indicates
that the effect of a single atom is small. An ensemble with $10^{3}-10^{4}$
atoms will show significant effects.}
\end{figure}

\begin{figure}
\includegraphics[width=15cm,height=8cm]{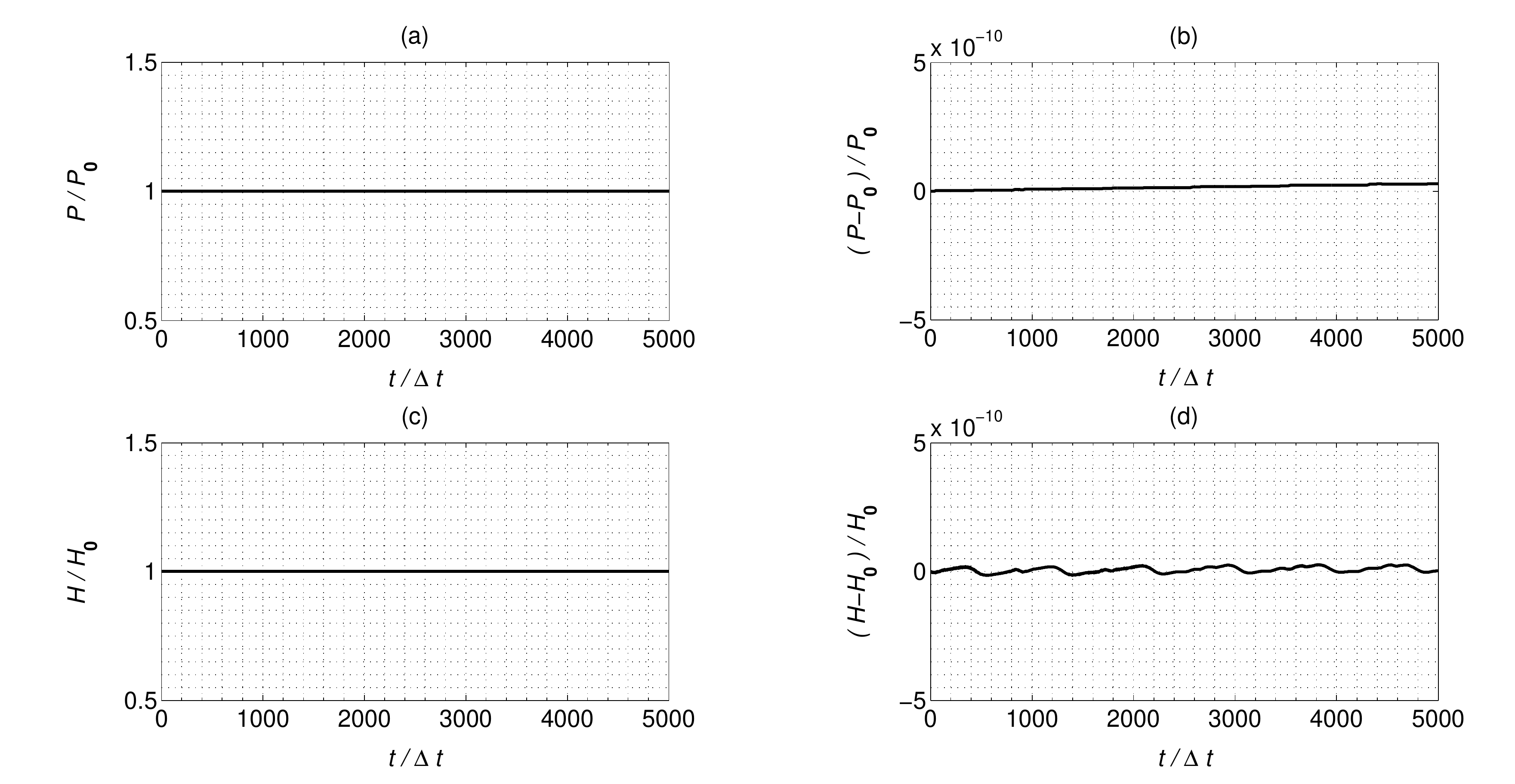} \caption{\label{fig:7} Time-history of numerical errors. After a long-term
simulation, both total probability error (a)-(b) and total Hamiltonian
error (c)-(d) are bounded by a small value.}
\end{figure}

In the second example, we simulate the continuous ionization of a
hydrogen atom in an ultrashort intense pulse-train of electromagnetic
field. Because the light-electron speed ratio is about 137, the coupling
between a single ultrashort pulse and the atom is weak. But with the
continuous excitation by the intense pulse-train, the atom can be
ionized gradually. The computation domain and initial wavefunction
are the same as the first example, and the time step is chosen to
$\bigtriangleup{t}=0.1\delta/\sqrt{3}c$ a.u. to capture the scattering
process. To introduce the incident pulse-train, we set the initial
gauge field to be $\bm{A}^{0}=100e^{-(z+2.5)^{2}/0.25}\bm{e}_{x}$
and $\bm{Y}^{0}=0$, representing two linearly-polarized
modulated Gaussian waves which counter-propagate along the
$z$-direction. The evolution of wave function is plotted in Figs.~\ref{fig:4}
and \ref{fig:5}, which depicts the continuous ionization process
by the ultrashort intense pulse-train. The ionization is indicated
by the increasing plane-wave components of the wavefunction. Figure~\ref{fig:6}
illustrates the evolution of scattered gauge field, which dependents
strongly on the electron polarization current. 
To demonstrate the excellent long-term properties of the structure-preserving
geometric algorithm, the time-history of numerical errors in this
example are plotted in Fig.~\ref{fig:7}. After a long-term simulation,
the numerical errors of conservation quantities are bounded by a small
value. 

\section{Conclusions}

\label{sec:6}

The structure-preserving geometric algorithms developed provide us
with a first-principle based simulation capability for the SM system
with long-term accuracy and fidelity. Two numerical examples validated
the algorithm and demonstrated its applications. This approach is particularly valuable when the
laser intensity reaches $10^{18}$ W$\cdot$cm$^{-2}$, which invalidates
many reduced or simplified theoretical and numerical models based
on perturbative analysis. For example, structure-preserving geometric
algorithms can be applied to achieve high fidelity simulations of
the HHG physics and the stabilization effect of ionization. The HHG
has been partially explained by the three-step semi-classical model
and the Lewenstein model in the strong field approximation \cite{Krause1992,Corkum1993,Lewenstein1994}.
After ionization, acceleration and recapture in a strong field, the
electron emits photons with a high order harmonic spectrum. The step
and cutoff structures of the spectrum strongly depend on the beam
intensity, photon energy and atomic potential. With the time dependent
wave function, the spectrum $F(\omega)=\int_{T}\int_{V}\psi^{*}(t)\ddot{\bm{x}}\psi(t)e^{i\omega{t}}\mathrm{d}^{3}x\mathrm{d}t$
can be calculated numerically. It can also be obtained by calculating the scattered gauge field spectrum
via a class of numerical probes around the potential center. Numerically
calculated wave functions also contains detailed information about
the dynamics of ionization. In a strong field, the atomic potential
is seriously dressed, and the wave function becomes non-localized.
Therefore, electrons have a chance of jumping into free states. Above
a specified threshold, the stabilization will quickly appears, i.e.,
the ionization rate increases slowly with the growth of beam intensity
and photon energy \cite{Pont1990,Eberly1993}. By introducing a proper
absorbing boundary condition in the simulation, the ionization rate
can be calculated as $\Gamma_{I}=\oint\frac{1}{2}(\psi_{R}\bigtriangledown\psi_{I}-\psi_{I}\bigtriangledown\psi_{R}){\cdot}\mathrm{d}\bm{S}$,
which gives a non-perturbative numerical treatment of the phenomena.

\section*{Acknowledgements}

\label{ack}

This research is supported by the National Natural Science Foundation
of China (NSFC-51477182, 11575185, 11575186), ITER-China Program (2015GB111003)
and Key Research Program of Frontier Sciences CAS (QYZDB-SSW-SYS004).



\begin{thebibliography}{99}
\bibitem{Mourou2007} G.~A. Mourou, C.~L. Labaune, M.~Dunne, N.~Naumova,
V.~T. Tikhonchuk, Relativistic laser-matter interaction: from attosecond
pulse generation to fast ignition, Plasma Phys. Contr. Fusion 49 (2007)
B667.

\bibitem{Krausz2009} F.~Krausz, M.~Ivanov, Attosecond physics,
Rev. Mod. Phys. 81 (2009) 163.

\bibitem{Voronov1965} G.~Voronov, N.~Delone, Multiphoton ionization
of krypton and argon by ruby laser radiation, JEPT Lett. 1 (1965)
66.

\bibitem{Keldysh1965} L.~V. Keldysh, Ionization in the field of
a strong electromagnetic wave, Sov. Phys. JETP 20 (1965) 1307.

\bibitem{Faisal1973} F.~H.~M. Faisal, Multiple absorption of laser
photons by atoms, J. Phys. B 6 (1973) L89.

\bibitem{Agostini1979} P.~Agostini, F.~Fabre, G.~Mainfray, G.~Petite,
N.~K. Rahman, Free-free transitions following six-photon ionization
of xenon atoms, Phys. Rev. Lett. 42 (1979) 1127.

\bibitem{Reiss1980} H.~R. Reiss, Effect of an intense electromagnetic
field on a weakly bound system, Phys. Rev. A 22 (1980) 1786.

\bibitem{Gontier1980} Y.~Gontier, N.~K. Rahman, M.~Trahin, Multiphoton
absorptions above the ionization threshold, J. Phys. B 13 (1980) 1381.

\bibitem{McPherson1987} A.~McPherson, G.~Gibson, H.~Jara, U.~Johann,
T.~S. Luk, Studies of multiphoton production of vacuum-ultraviolet
radiation in the rare gases, J. Opt. Soc. Am. B 4 (1987) 595.

\bibitem{Gallagher1988} T.~F. Gallagher, Above-threshold ionization
in low-frequency limit, Phys. Rev. Lett. 61 (1988) 2304.

\bibitem{Eberly1991} J.~H. Eberly, J.~Javanainen, K.~Rzazewski,
Above-threshold ionization, Phys. Rep. 204 (1991) 331.

\bibitem{Pont1990} M.~Pont, M.~Gavrila, Stabilization of atom hydrogen
in superintense, high-frequency laser fields of circular polarization,
Phys. Rev. Lett. 65 (1990) 2362.

\bibitem{Krause1992} J.~L. Krause, K.~J. Schafer, K.~C. Kulander,
High-order harmonic generation from atoms and ions in the high intensity
regime, Phys. Rev. Lett. 68 (1992) 3535.

\bibitem{Eberly1993} J.~H. Eberly, K.~C. Kulander, Atomic stabilization
by super-intense lasers, Science 262 (1993) 5137.

\bibitem{Corkum1993} P.~Corkum, Plasma perspective on strong-field
multiphoton ionization, Phys. Rev. Lett. 71 (1993) 1994.

\bibitem{Lewenstein1994} M.~Lewenstein, P.~Balcou, M.~Ivanov,
A.~L'Huillier, P.~Corkum, Theory of high-harmonic generation by
low-frequency laser fields, Phys. Rev. A 49 (1994) 2117.

\bibitem{Birula1994} I.~Bialynicki-Birula, M.~Kali{\'{n}}ski,
J.~H. Eberly, Lagrange equilibrium points in celestial mechanics
and nonspreading wave packets for strongly driven Rydberg electrons,
Phys. Rev. Lett. 73 (1994) 1777.

\bibitem{Bao1996} D.~Bao, S.~G. Chen, J.~Liu, Rescattering effect
in above-threshold ionization processes, J. Appl. Phys. B 62 (1996)
313.

\bibitem{Spielmann1997} C.~Spielmann, N.~H. Burnett, S.~Sartania,
R.~Koppitsch, M.~Schn\"{u}rer, C.~Kan, M.~Lenzner, P.~Wobrauschek,
F.~Krausz, Generation of coherent X-rays in the water window using
5-femtosecond laser pulses, Science 278 (1997) 661.

\bibitem{Sali2001} P.~Sali{\`{e}}res, B.~Carr{\'{e}}, L.~L.
D{\`{e}}roff, F.~Grasbon, G.~G. Paulus, H.~Walther, R.~Kopold,
W.~Becker, D.~B. Milo{\u{s}}evi{\'{c}}, A.~Sanpera, M.~Lewenstein,
Feynman's path-integral approach for intense-laser-atom interactions,
Science 292 (2001) 902.

\bibitem{Kienberger2004} R.~Kienberger, E.~Goulielmakis, M.~Uiberacker,
A.~Baltuska, V.~Yakovlev, F.~Bammer, A.~Scrinzi, T.~Westerwalbesloh,
U.~Kleineberg, U.~Heinzmann, M.~Drescher, F.~Krausz, Atomic transient
recorder, Nature 427 (2004) 817.

\bibitem{Drake2006} R.~P. Drake, High Energy Density Physics: Fundamentals,
Inertial Fusion and Experimental Astrophysics, Springer, New York,
2006.

\bibitem{Brabec2008} T.~Brabec, H.~Kapteyn, Strong Field Laser
Physics, Springer, New York, 2008.

\bibitem{Kim2008} S.~Kim, J.~Jin, Y.-J. Kim, I.-Y. Park, Y.~Kim,
S.-W. Kim, High-harmonic generation by resonant plasmon field enhancement,
Nature 453 (2008) 757.

\bibitem{Smirnova2009} O.~Smirnova, Y.~Mairesse, S.~Patchkovskii,
N.~Dudovich, D.~Villeneuve, P.~Corkum, M.~Y. Ivanov, High harmonic
interferometry of multi-electron dynamics in molecules, Nature 460
(2009) 972.

\bibitem{Le2009} A.-T. Le, R.~R. Lucchese, S.~Tonzani, T.~Morishita,
C.~D. Lin, Quantitative rescattering theory for high-order harmonic
generation from molecules, Phys. Rev. A 80 (2009) 013401.

\bibitem{Goulielmakis2010} E.~Goulielmakis, Z.-H. Loh, A.~Wirth,
R.~Santra, N.~Rohringer, V.~S. Yakovlev, S.~Zherebtsov, T.~Pfeifer,
A.~M. Azzeer, M.~F. Kling, S.~R. Leone, F.~Krausz, Real-time observation
of valence electron motion, Nature 466 (2010) 739.

\bibitem{Nepstad2010} R.~Nepstad, T.~Birkeland, M.~F{\o }rre,
Numerical study of two-photon ionization of helium using an ab initio
numerical framework, Phys. Rev. A 81 (2010) 063402.

\bibitem{Birkeland2010} T.~Birkeland, R.~Nepstad, M.~F{\o }rre,
Stabilization of helium in intense XUV laser fields, Phys. Rev. Lett.
104 (2010) 163002.

\bibitem{Popmintchev2012} T.~Popmintchev, M.-C. Chen, D.~Popmintchev,
P.~Arpin, S.~Brown, S.~Ali{\u{s}}auskas, G.~Andriukaitis, T.~Bal{\u{c}}iunas,
O.~D. M\"{u}cke, A.~Pugzlys, A.~Baltu{\u{s}}ka, B.~Shim, S.~E.
Schrauth, A.~Gaeta, C.~Hern{\'{a}}ndez-Garc{\'{i}}a, L.~Plaja,
A.~Becker, A.~Jaron-Becker, M.~M. Murnane, H.~C. Kapteyn, Bright
coherent ultrahigh harmonics in the KeV X-ray regime from mid-infrared
femtosecond lasers, Science 336 (2012) 1287.

\bibitem{Piazza2012} A.~D. Piazza, C.~M\"{u}ller, K.~Z. Hatsagortsyan,
C.~H. Keitel, Extremely high-intensity laser interactions with fundamental
quantum systems, Rev. Mod. Phys. 84 (2012) 1177.

\bibitem{Becker2012} W.~Becker, X.~Liu, P.~J. Ho, J.~H. Eberly,
Theories of photoelectron correlation in laser driven multiple atomic
ionization, Rev. Mod. Phys. 84 (2012) 1011.

\bibitem{Madsen2012} C.~B. Madsen, F.~Anis, L.~B. Madsen, B.~D.
Esry, Multiphoton above threshold effects in strong-field fragmentation,
Phys. Rev. Lett. 109 (2012) 163003.

\bibitem{Argenti2013} L.~Argenti, R.~Pazourek, J.~Feist, S.~Nagele,
M.~Liertzer, E.~Persson, J.~Burgd\"{o}rfer, E.~Lindroth, Photoionization
of helium by attosecond pulses: extraction of spectra from correlated
wave functions, Phys. Rev. A 87 (2013) 053405.

\bibitem{Yuan2013} K.-J. Yuan, A.~D. Bandrauk, Single circularly
polarized attosecond pulse generation by intense few cycle elliptically
polarized laser pulses and terahertz fields from molecular media,
Phys. Rev. Lett. 110 (2013) 023003.

\bibitem{Guo2013} L.~Guo, S.~S. Han, X.~Liu, Y.~Cheng, Z.~Z.
Xu, J.~Fan, J.~Chen, S.~G. Chen, W.~Becker, C.~I. Blaga, A.~D.
DiChiara, E.~Sistrunk, P.~Agostini, L.~F. DiMauro, Scaling of the
low-energy structure in above-threshold ionization in the tunneling
regime: theory and experiment, Phys. Rev. Lett. 110 (2013) 013001.

\bibitem{Klaiber2013} M.~Klaiber, E.~Yakaboylu, K.~Z. Hatsagortsyan,
Above-threshold ionization with highly charged ions in superstrong
laser fields. II. relativistic Coulomb-corrected strong-field approximation,
Phys. Rev. A 87 (2013) 023418.

\bibitem{Vampa2014} G.~Vampa, C.~McDonald, G.~Orlando, D.~Klug,
P.~Corkum, T.~Brabec, Theoretical analysis of high-harmonic generation
in solids, Phys. Rev. Lett. 113 (2014) 073901.

\bibitem{Popruzhenko2014} S.~V. Popruzhenko, Keldysh theory of strong
field ionization: history, applications, difficulties and perspectives,
J. Phys. B 47 (2014) 204001.

\bibitem{Popmintchev2015} D.~Popmintchev, C.~Hern{\'{a}}ndez-Garc{\'{i}}a,
F.~Dollar, C.~Mancuso, J.~A. P{\'{e}}rez-Hern{\'{a}}ndez, M.-C.
Chen, A.~Hankla, X.~Gao, B.~Shim, A.~L. Gaeta, M.~Tarazkar, D.~A.
Romanov, R.~J. Levis, J.~A. Gaffney, M.~Foord, S.~B. Libby, A.~Jaron-Becker,
A.~Becker, L.~Plaja, M.~M. Murnane, H.~C. Kapteyn, T.~Popmintchev,
Ultraviolet surprise: efficient soft X-ray high-harmonic generation
in multiply ionized plasmas, Science 350 (2015) 1225.

\bibitem{Kfir2015} O.~Kfir, P.~Grychtol, E.~Turgut, R.~Knut,
D.~Zusin, D.~Popmintchev, T.~Popmintchev, H.~Nembach, J.~M. Shaw,
A.~Fleischer, H.~Kapteyn, M.~Murnane, O.~Cohen, Generation of
bright phase-matched circularly-polarized extreme ultraviolet high
harmonics, Nature Photon. 9 (2015) 99.

\bibitem{Luu2015} T.~T. Luu, M.~Garg, S.~Y. Kruchinin, A.~Moulet,
M.~T. Hassan, E.~Goulielmakis, Extreme ultraviolet high-harmonic
spectroscopy of solids, Nature 521 (2015) 498.

\bibitem{Bukov2015} M.~Bukov, L.~D'Alessio, A.~Polkovnikov, Universal
high-frequency behavior of periodically driven systems: from dynamical
stabilization to Floquet engineering, Adv. Phys. 64 (2015) 139.

\bibitem{Hassan2016} M.~T. Hassan, T.~T. Luu, A.~Moulet, O.~Raskazovskaya,
P.~Zhokhov, M.~Garg, N.~Karpowicz, A.~M. Zheltikov, V.~Pervak,
F.~Krausz, E.~Goulielmakis, Optical attosecond pulses and tracking
the nonlinear response of bound electrons, Nature 530 (2016) 66.

\bibitem{Corkum11} P.~Corkum, Recollision physics, Physics Today
64 (2011) 36.

\bibitem{Hochstuhl2014} D.~Hochstuhl, C.~Hinz, M.~Bonitz, Time-dependent
multiconfiguration methods for the numerical simulation of photoionization
processes of many-electron atoms, Eur. Phys. J. Spec. Top. 223 (2014)
177.

\bibitem{Miyagi2014} H.~Miyagi, L.~B. Madsen, Time-dependent restricted-active-space
self-consistent-field theory for laser-driven many-electron dynamics.
II. extended formulation and numerical analysis, Phys. Rev. A 89 (2014)
063416.

\bibitem{Bauch2014} S.~Bauch, L.~K. S{\o }rensen, L.~B. Madsen,
Time-dependent generalized-active-space configuration-interaction
approach to photoionization dynamics of atoms and molecules, Phys.
Rev. A 90 (2014) 062508.

\bibitem{Wu1995} L.~Wu, X.~Jin, Z.~Wu, Symplectic structure of
Schr\"{o}dinger equation and symplectic algorithms for quantum mechanics,
Chin. J. comput. phys. 12 (1995) 127.

\bibitem{Yee1966} K.~S. Yee, Numerical solution of initial boundary
value problems involving Maxwell's equations in isotropic media, IEEE
Trans. Antenn. Propag. 14 (1966) 302.

\bibitem{Mur1981} G.~Mur, Absorbing boundary conditions for the
finite-difference approximation of the time-domain electromagnetic-field
equations, IEEE Trans. Electromagn. Compat. 23 (1981) 377.

\bibitem{Berenger1994} J.~P. Berenger, A perfectly matched layer
for the absorption of electromagnetic waves, J. comput. phys. 114
(1994) 185.

\bibitem{Blanes2006} S.~Blanes, F.~Casas, A.~Murua, Symplectic
splitting operator methods for the time-dependent Schr\"{o}dinger
equation, J. Chem. Phys. 124 (2006) 234105.

\bibitem{Kormann2008} K.~Kormann, S.~Holmgren, H.~O. Karlsson,
Accurate time propagation for the Schr\"{o}dinger equation with an
explicitly time-dependent Hamiltonian, J. Chem. Phys. 128 (2008) 184101.

\bibitem{Shen2013} J.~Shen, Wei E.~I. Sha, Z.~Huang, M.~Chen, X.~Wu,
High-order symplectic FDTD scheme for solving a time-dependent Schr\"{o}dinger
equation, Comput. Phys. Comm. 184 (2013) 445.

\bibitem{Squire12} J.~Squire, H.~Qin, W.~M. Tang, Geometric integration
of the {V}lasov-{M}axwell system with a variational particle-in-cell
scheme, Phys. Plasmas 19 (2012) 084501.

\bibitem{JXiao2013} J.~Xiao, J.~Liu, H.~Qin, Z.~Yu, A variational
multi-symplectic particle-in-cell algorithm with smoothing functions
for the Vlasov-Maxwell system, Phys. Plasmas 20 (2013) 102517.

\bibitem{JXiao2015} J.~Xiao, J.~Liu, H.~Qin, Z.~Yu, N.~Xiang,
Variational symplectic particle-in-cell simulation of nonlinear mode
conversion from extraordinary waves to Bernstein waves, Phys. Plasmas
22 (2015) 092305.

\bibitem{Xiao15-112504} J.~Xiao, H.~Qin, J.~Liu, Y.~He, R.~Zhang,
Y.~Sun, Explicit high-order non-canonical symplectic particle-in-cell
algorithms for Vlasov-Maxwell systems, Phys. Plasmas 22 (2015) 112504.

\bibitem{He15-124503} Y.~He, H.~Qin, Y.~Sun, J.~Xiao, R.~Zhang,
J.~Liu, Hamiltonian integration methods for Vlasov-Maxwell equations,
Phys. Plasmas 22 (2015) 124503.

\bibitem{QHong2016} H.~Qin, J.~Liu, J.~Xiao, R.~Zhang, Y.~He,
Y.~Wang, Y.~Sun, J.~W. Burby, L.~Ellison, Y.~Zhou, Canonical
symplectic particle-in-cell method for long-term large-scale simulations
of the Vlasov-Maxwell equations, Nucl. Fusion 56 (2016) 014001.

\bibitem{He16-092108} Y.~He, Y.~Sun, H.~Qin, J.~Liu, Hamiltonian
particle-in-cell methods for Vlasov-Maxwell equations, Phys. Plasmas
23 (2016) 092108.

\bibitem{Morrison2017} P.~Morrison, Structure and structure-preserving
algorithms for plasma physics, Phys. Plasmas 24 (2017) 055502.

\bibitem{Xiao-M2016} J.~Xiao, H.~Qin, P.~Morrison, J.~Liu, Z.~Yu,
R.~Zhang, Y.~He, Explicit high-order noncanonical symplectic algorithms
for ideal two-fluid systems, Phys. Plasmas 23 (2016) 112107.

\bibitem{Michael-ar} M.~Kraus, K.~Kormann, P.~Morrison, E.~Sonnendr\"{u}cker,
GEMPIC: Geometric ElectroMagnetic Particle-In-Cell Methods, https://arxiv.org/abs/1609.03053

\bibitem{Masiello2004} D.~Masiello, On the canonical formulation
of electromagnetics and wave mechanics, Ph.D. thesis, University of
Florida (2004).

\bibitem{Masiello2005} D.~Masiello, E.~Deumens, Y.~\"{O}hrn, Dynamics
of an atomic electron and its electromagnetic field in a cavity, Phys.
Rev. A 71 (2005) 032108.

\bibitem{Landau1965} L.~D. Landau, E.~M. Lifshitz, Quantum Mechanics:
Nonrelativistic Theory, Pergamon Press, Oxford, 1965.

\bibitem{Dirac1958} P.~A.~M. Dirac, The Principles of Quantum Mechanics
(4th ed.), Oxford University Press, Oxford, 1958.
\end{thebibliography}


\section*{References}

\providecommand{\noopsort}[1]{}\providecommand{\singleletter}[1]{#1}

\expandafter\ifx\csname url\endcsname\relax \global\long\def\url#1{\texttt{#1}}
\fi \expandafter\ifx\csname urlprefix\endcsname\relax\global\long\def\urlprefix{URL }
\fi \expandafter\ifx\csname href\endcsname\relax \global\long\def\href#1#2{#2}
 \global\long\def\path#1{#1}
\fi

\end{document}